\documentclass[aps,pre,twocolumn,superscriptaddress,showpacs,preprintnumbers]{revtex4-2}
\usepackage{natbib}
\usepackage[english]{babel}
\usepackage[dvips]{graphics}
\usepackage{graphicx,epsfig}
\usepackage{amsmath}
\usepackage{color}
\usepackage{multirow}
\usepackage[normalem]{ulem}
\usepackage{booktabs}
\usepackage{physics}
\usepackage{dsfont}
\usepackage{amssymb}
\usepackage{lineno}
\begin{document}
\title{Using reservoir computing to construct scarred wavefunctions}
%
\author{L. Domingo}
\email[E--mail address: ]{laia.domingo@icmat.es}
\affiliation{Departamento de Química; Universidad Autónoma de Madrid;
Cantoblanco - 28049 Madrid, Spain}
\affiliation{Grupo de Sistemas Complejos; Universidad Politécnica de Madrid; 28035 Madrid, Spain}
\affiliation{Instituto de Ciencias Matemáticas (ICMAT); Campus de Cantoblanco; 
Nicolás Cabrera, 13-15; 28049 Madrid, Spain}
\author{J. Borondo}
\email[E--mail address: ]{jborondo@gmail.com}
\affiliation{Departamento de Gestión Empresarial; Universidad Pontificia de Comillas; Madrid; Spain}
\affiliation{AgrowingData, Almería, Spain}
\author{F. Borondo}
\email[E--mail address: ]{f.borondo@uam.es}
\affiliation{Departamento de Química; Universidad Autónoma de Madrid;
Cantoblanco - 28049 Madrid, Spain}

\date{\today}
\begin{abstract}
Scar theory is one of the fundamental pillars in the field of quantum chaos, 
and scarred functions a superb tool to carry out studies in it.
Several methods, usually semiclassical, have been described to cope with these two phenomena.
In this paper, we present an alternative method, based on the novel machine
learning algorithm known as Reservoir Computing, to calculate such scarred wavefunctions 
together with the associated eigenstates of the system.
The resulting methodology achieves outstanding accuracy while reducing execution times by a factor of ten.
As an illustration of the effectiveness of this method, we apply it to the widespread chaotic 
two-dimensional coupled quartic oscillator.
\end{abstract}

\maketitle

\section{Introduction}
\label{sec:Intro}

At the end of the last century, significant attention was paid to the correspondence 
between classical and quantum dynamics, particularly in scenarios in which classical chaos dominates \cite{Gutzwiller}.
Over time, this evolution gave rise to the field of quantum chaos, which explores how the principles 
of quantum mechanics apply to classically chaotic systems, or how classical chaos
manifests in the quantum world \cite{Berr89}.
Two main pillars in the field emerged.
First, Bohigas, Giannoni and Schmit \cite{bohigas} found that the level fluctuations of the quantum 
Sinai's billiard is consistent with the predictions of the Gaussian orthogonal ensemble of random 
matrices, thus reinforcing the belief that level fluctuation laws are universal.
Second, Heller \cite{Heller} coined the term scar to describe the increase of the probability density 
that some eigenfunctions exhibit along unstable periodic orbits (POs) of a classically chaotic system. 
This property came as a surprise since, even though the chaotic classical dynamics cause the classical trajectories to deviate from the POs, many eigenfunctions tend to concentrate near those POs.
Later, this definition \cite{Kaplan} was generalized resulting in the so-called scarred functions
\cite{Pol94,Ver01}, which are not only localized along the PO in configuration space, 
but also along its invariant manifolds in the phase space \cite{Fabio1}. 
More recently, other interesting perspectives were brought up on the issue. 
Among them, the use of Loschmidt echoes or fidelity functions \cite{Lecho}, out-of-time-order correlators 
\cite{OTOC}, or Krylov operators \cite{Krylov} deserve special consideration. 

The scarring phenomenon  plays a central role in the study of the correspondence between classical 
and quantum mechanics in the presence of chaos. 
For example, scar functions have been used to understand the properties of eigenstates in 
quantum chaotic systems, since they form an efficient basis set for calculation
of the eigenstates of such systems \cite{Kaplan,Wis3,Wis4,Fabio1,Hummel}.

In the field of machine learning, many algorithms have been designed to predict the time evolution 
of dynamical systems \cite{RNN, LSTM}. 
A particularly relevant algorithm for this task is called Reservoir Computing (RC),
which has proven very useful in chaotic time series prediction \cite{chaosRC1, chaosRC2,chaosRC3}. 
The underlying idea is that a well-trained reservoir can reproduce the attractor 
of the originating chaotic dynamics.  
Apart from its excellent performance in predicting the input-output dynamics, 
because of its simple training strategy, it is usually computationally more efficient 
than other popular machine learning algorithms. 
In a previous work \cite{Domingo}, we provided a method to adapt the classical RC algorithm to
perform the time evolution of a quantum wavefunction in time, 
under the effect of a certain Hamiltonian. The performance of this method was illustrated by integrating the time-dependent Schrödinger equation 
for some simple 1D and 2D quantum systems, such as the harmonic oscillator and the Morse Hamiltonian. 

In this work, we apply this method to a more complex and realistic quantum system, 
the coupled quartic oscillator, which has been of great interest in the field of quantum chaos
\cite{Fabio1,Fabio2,chaotic}. 
The key idea is to start from a suitable initial wavepacket at a certain energy, 
and the use of RC to propagate this state in time. 
From the time evolution of the wavefunction, we can obtain the energy spectrum and the eigenstates 
of the system around the same energy as the initial state. 
This method also allows us to obtain the scarred functions of the quantum chaotic system. 
Similar studies have been recently reported \cite{HenonHeiles} using the Hénon-Heiles 2D potential 
as a model for molecular vibrations.

The organization of this paper is as follows. 
Section~\ref{sec:methods} describes the coupled quartic oscillator, the quantum system studied in this work,
and the methods used in our study. 
In particular, Sect.~\ref{RC} presents the original formalism of RC, and in 
Sect.~\ref{sec:RC_QuantumSystems} we describe the adaptations that have to be made to integrate 
the Schrödinger equation. 
The corresponding results are presented in Sect.~\ref{sec:Results}.  
Finally, Sect. \ref{sec:Conclusions} ends the paper by summarizing the main conclusions of the present work.

\section{System and methods}
\label{sec:methods}
\subsection{System: the two-dimensional coupled quartic oscillator}
\label{quartic}
The system that we have chosen to study is the dynamics of a unit mass particle moving in a two-dimensional
coupled quartic potential described with the following Hamiltonian
\begin{equation}
    H(x,y,p_x,p_y) = \frac{1}{2}\left(p_x^2 + p_y^2\right) + \frac{1}{2} \; x^2y^2 
             + \frac{\epsilon}{4} \, \left(x^4 + y^4\right),
    \label{eq:quartic}
\end{equation}
with $\epsilon=0.01$. 
The potential of this system is analytic and induces very chaotic dynamics \cite{chaotic,BohigasPR}, 
reasons why this system has been extensively studied in the field of quantum chaos
\cite{Fabio1,Fabio2,Fabio3,quartic1,quartic2,BohigasPR}. 
Moreover, this potential, which belongs to the $C_{4v}$ symmetry group, is homogeneous, 
and then the classical trajectories exhibit mechanical similarity.
That is, any trajectory  $(x(t), y(t), p_x(t), p_y(t))$ run at a certain energy E can be scaled to
$(x'(t), y'(t), p'_x(t), p'_y(t))$ at energy $E'$, by using the following scaling relations,
\begin{equation}
        \left\{\begin{array}{ll}
        & x' := \eta x, \quad p_x' := \eta^2 p_x, \\
        & y' := \eta y, \quad p_y' := \eta^2 p_y, \\
        & t' := t/\eta, \quad S'    = \eta^3 S,
    \end{array}\right.
    \label{eq:scaling}
\end{equation}
where $\eta =  \displaystyle \left(\frac{E'}{E} \right)^{1/4}$, and being $S$ the corresponding classical 
action $S = \int_0^T (p_x \dot{x} + p_y \dot{y})\,dt$.

\subsection{Periodic orbits}

PO trajectories for the quartic potential can be calculated by solving the system of 
differential equations
    \begin{equation}
        \left\{\begin{array}{ll}
        &\dot{x} = p_x,  \\
        &\dot{y} = p_y,  \\
        &\dot{p}_x = -(xy^2 + \epsilon x^3), \\
        &\dot{p}_y = -(x^2y + \epsilon y^3),  \\
        &\dot{S}_x = p_x\dot{x} = p^2_x, \\
        &\dot{S}_y = p_y\dot{y} = p^2_y,
    \end{array}\right.
    \end{equation}
where the appropriate initial conditions are chosen to obtain the desired PO.
These can be obtained by taking into account the (high) symmetry of the system \cite{TesisFabio}.
Notice that the last two equations are added to calculate the classical action, 
which is also needed in our calculations. 

\begin{figure}
 \includegraphics[width=1.0\columnwidth]{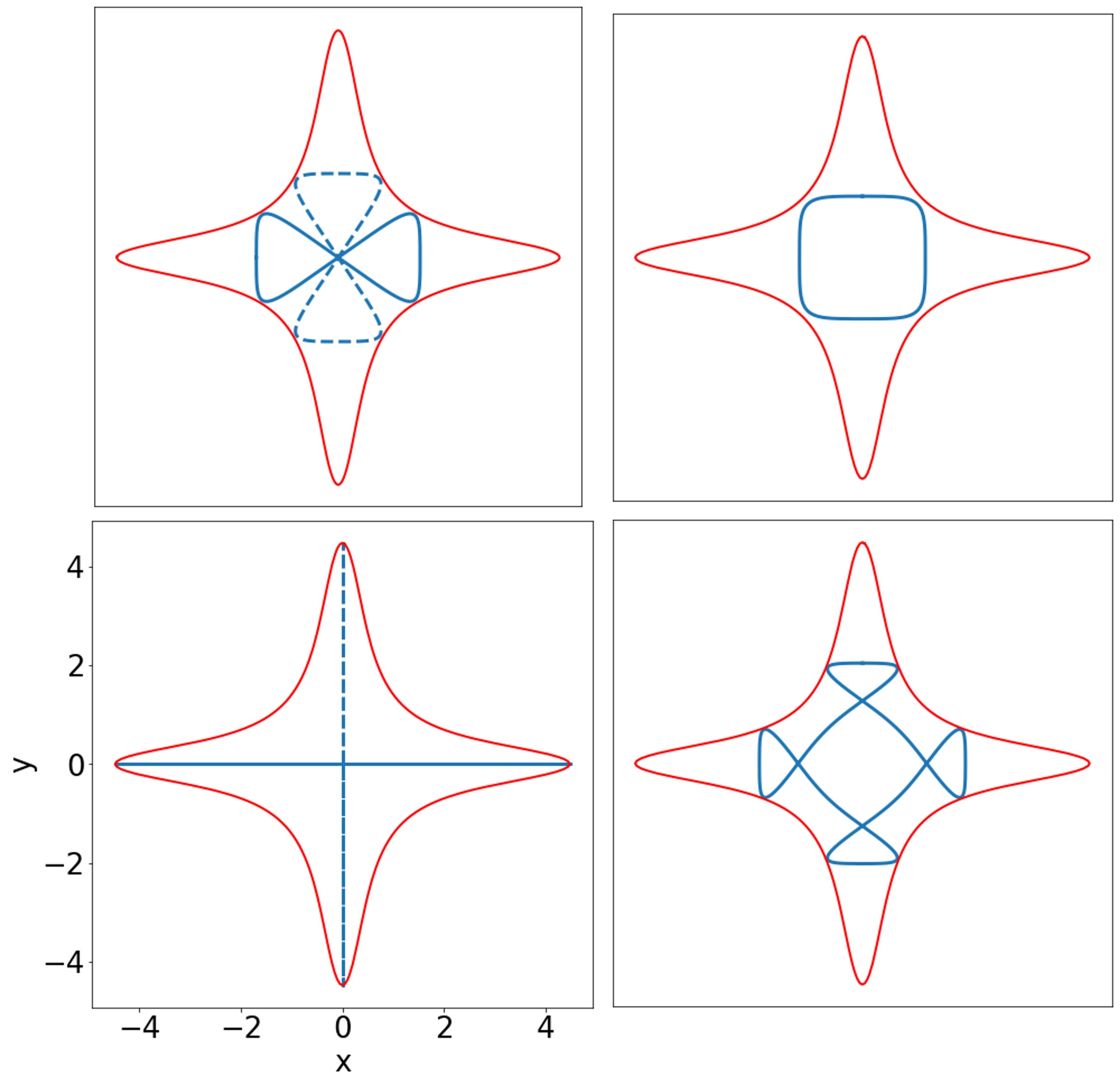}
  \caption{Trajectories corresponding to the four periodic orbits in blue (light gray):
  (from bottom left to upper right) 
  horizontal/vertical, quadruple-loop, horizontal/vertical bowtie, and square, at $E=1$.
  The corresponding equipotential is shown with the red (dark gray) line.}
 \label{fig:po}
\end{figure}

In Fig.~\ref{fig:po}, we show the four POs considered in this work,
which correspond to the horizontal/vertical, quadruple-loop, horizontal/vertical bowtie, 
and square trajectories of our system, respectively.

\subsection{Quantum dynamics}
\label{sect:qdynamics}

The quantum dynamics of a system is obtained by solving the time-dependent Schrödinger equation, 
\begin{equation}
     i \hbar \; \frac{\partial \psi(x,y,t)}{\partial t} = \hat{H} \psi(x,y,t),
    \label{eq:Schrodinger}
\end{equation}
where $\hat{H}$ is the associated Hamiltonian operator, which in our case was discussed in the 
previous Sect.~\ref{quartic}. 
This partial differential equation can be integrated very efficiently using the fast Fourier transform
(FFT) method developed by Kosloff and Kosloff \cite{kosloff}.
With this method, an initial quantum state $\psi_0(x,y)$, 
with certain (mean) energy $E_0$, can be propagated in time obtaining the corresponding time evolved 
state $\psi(x,y,t)$. 

This constitutes the foundation of our computations, in which we will use the method described 
above to train the RC approach described in Sect.~\ref{sec:RC_QuantumSystems}.

\subsection{Eigenfunctions}
\label{sect:eigenfunctions}
From the time evolved $\psi(x,y,t)$ of a suitable initial quantum state $\psi_0(x,y)$ 
the eigenenergies and eigenfunctions of the quantum system with similar energies to $E_0$ 
can be calculated, by performing the following steps
\begin{enumerate}
    \item Prepare an initial wavefunction $\psi_0(x,y)$ as a minimum uncertainty Gaussian wave packet 
    with mean energy $E_0$.
    \begin{eqnarray}
        \psi_0(x,y) & = & \displaystyle \pi^{-1/8} \; e^{-(x-x_0)^2/(4 \Delta x)^2} \nonumber \\
        & & \times \, e^{-(y-y_0)^2/(4 \Delta y)^2} \nonumber \\
        & & \times \, e^{i\,(p_{x,0}x + p_{y,0} y)/\hbar},
        \label{eq:psi0}
    \end{eqnarray}
    where $(x_0, y_0)$ is the central position, $\Delta x, \Delta y$ are the spread in the $x$ and $y$ directions, and $(p_{x,0}, p_{y,0})$ is the momentum of the wavepacket. 
    The energy of the wavepacket in Eq.~(\ref{eq:psi0}) corresponds to the value of the Hamiltonian 
    function (\ref{eq:quartic}), at the central position $(x_0, y_0)$ and momentum $(p_{x,0}, p_{y,0})$.
    \item Compute the time evolution $\psi(x,y,t)$ under Hamiltonian $\hat{H}(x,y)$. 
    For the first $t_\text{train}$ time steps we use the FFT method proposed by 
    Kosloff and Kosloff \cite{kosloff}. 
    For the next $t_\text{test}$ steps we use the RC algorithm proposed in 
    Ref.~\cite{Domingo} and summarized in Sect.~\ref{RC} below.
    \item Calculate the energy spectrum $I(E)$. 
    For this purpose, we first calculate the time-correlation function 
    \begin{eqnarray}
        C(t) = \braket{\psi(t)}{\psi(0)} =\\ \nonumber \int_{(x,y)} \; \psi(x,y,t)^* \psi_0(x,y) \; dx\,dy.
        \label{eq:time-corr}
    \end{eqnarray}

Then, we obtain the energy spectrum $I(E)$ as the Fourier transform of the time-correlation function
    \begin{equation}
        I(E) = \int_t \; C(t) \; e^{-i E t/ \hbar} \; dt ,
        \label{eq:spectrum}
    \end{equation}
    where the integral should be done for infinite time. 
    For practical purposes, however, the integral is done for a large interval of time $[0, T]$. 
    The eigenenergies $E_n$ correspond to the peaks of the energy spectrum function $|I(E)|$. 
    
    \item Calculate the eigenfunctions $\phi_n(x,y)$ by performing the Fourier transform of
    $\psi(x,y,t)$ at the eigenenergies $E_n$
    \begin{equation}
        \phi_n(x,y) = \int_t \; \psi(x,y,t) \; e^{i E_n t/\hbar} \; dt.
        \label{eq:eigenfunct}
    \end{equation}
Therefore, the problem is reduced to propagating an initial wavepacket in time by solving the time-dependent 
Schrödinger equation (in step 2) using RC. 
\end{enumerate}

In this work, we calculate the eigenfunctions and eigenenergies with $A_1$ symmetry, i.e.~most symmetric,
for three different initial conditions at energies $E_0=1, 10$ and $100$, respectively.
For simplicity of visualization, all the systems are scaled to energy $E=1$ using the scaling relations 
in Eq.~(\ref{eq:scaling}). 
The parameters of the initial wavepackets are reported in Table~\ref{tab:initial_waves}. 

At this point, two somewhat related issue should be considered: symmetry and propagation time.
Regarding the former,
notice that the symmetry of the initial wavepacket determines the set of eigenfunctions 
obtained with this method.
That is, if the initial wavepacket in Eq.~(\ref{eq:psi0}) has a certain symmetry 
(for example along the $x$ axis), the resulting wavefunctions will also have this same symmetry. 
Accordingly, if the whole set of eigenfunctions in a given energy interval wants to be collected, 
the previous steps need to be repeated with different symmetry-adapted $\psi_0(x,y)$'s 
taking into account all possible symmetries in the system.
Regarding the latter, consider that among that the eigenvalues resolution of our calculated is 
mainly dictated by the propagation time of the initial wavepacket. 
In the ideal case of an infinite time span, all eigenvalues would be resolved no matter how close 
they would be. 
However, this is not feasible in an actual numerical calculation, and then the propagation time 
has to be selected taking into account the density of states of the system. 
Fortunately, this function function can be semiclassically ascertained. 
The leading term in the corresponding expansion in $\hbar^{-1}$ is obtained by deriving the Weyl formulae, 
and higher order terms can also been added in the expansion, for example to account for the effect 
of the parity in the eigenfunctions existing along the different symmetry lines. 
This is mandatory in our case of the quartic oscillator, due to the high symmetry of the potential. 
Full details and explicit expressions for the quartic potential in our work can be found in 
Refs.~\cite{TesisFabio,Fabio2024}.

Moreover, also note that the parameter $t_\text{train}$ is influenced by the system initial energy, 
with higher energies requiring more training steps for the reservoir to effectively learn the system dynamics.
As system complexity rises with energy, enlarging $t_\text{train}$ becomes necessary due to the increased
challenge for the reservoir to capture intricate dynamics. 
Proper training allows the reservoir to autonomously predict future system dynamics. 
Consequently, a smaller $t_\text{train}$ could compromise the capacity of generalization of the method. 
Increasing $t_\text{train}$ to the values in Table~\ref{tab:initial_waves} ensures a correct training, 
reducing the risk of overfitting to the training dynamics.

\begin{table}
    \centering
    \begin{tabular}{rcccccccc}
    \hline \hline
      $E$ & $x_0$ & $y_0$ & $\Delta x$ & $\Delta y$  & $p_0$ & $t_\text{train}$ & $t_\text{test}$ & $\Delta t$\\
      \hline
       1   &  0    & 0     & 0.5        & 0.5        & 1    & 1500 &  4650 & 0.00140  \\
       10  & 1.860 & 1.860 & $2^{-1/2}$ & $2^{-1/2}$ & $-2$ & 3000 & 10000 & 0.00030  \\
       100 & 3.665 & 3.665 & 1          & 1          & $-3$ & 5000 &  8000 & 0.00003 \\
       \hline
    \end{tabular}
    \caption{Parameters of the initial wavefunctions, as described in Eq.~(\ref{eq:initial}) 
    at three different energies used in our calculations. 
    Training and test time for the RC algorithm, 
    and time intervals between steps are also given in the last three columns.}
    \label{tab:initial_waves}
\end{table}

\subsection{Scar functions}
\label{sec:scarfunctions}
    \begin{table*}
        \centering
        \begin{tabular}{lcccccc}
            \hline \hline
            Scar & $T$ & $ND$ & $P$ & $n$ & $\mu$ & $(x_0,y_0,p_{x_0}, p_{y_0})$ \\
            \hline    
            \multirow{2}{*}{Horizontal/vertical} & \multirow{2}{*}{33.17}  & \multirow{2}{*}{0} & \multirow{2}{*}{0} & 12,14,16,18,20,22, & \multirow{2}{*}{16} & \multirow{2}{*}{(0,0, $\sqrt{2}$,$\sqrt{2}$)} \\
             & & & & 42,47,53, 58,62 &  &\\
            Quadruple-loop  & 18.75 & 0 & 4 & 4,5,6,7,8,9,10 & 12 & (0, 2.028, 1.384, 0) \\
            Bowtie hor/ver  & 9.54  & 0 & 2 & 20,22,24       & 4  & (0, 1.655, 1.401, 0) \\
            Square          & 7.84  & 0 & 4 & 3,4,5,6,7,8    & 4  & (0, 1.239, 1.410, 0) \\
            \hline
        \end{tabular}
        \caption{Number of Dirichlet conditions on the wave functions at symmetry lines (ND), 
        ratio between the period of the full periodic orbit and that of the desymmetrized periodic orbit (P),
        excitation number (n), Maslov index ($\mu$) and initial conditions for the four periodic orbits 
        studied in this work, at energy $E=1$. 
        The values of all the parameters are taken from Ref.~\cite{TesisFabio}.}
        \label{tab:PO}
    \end{table*}
Apart from the eigenfunctions, we also use the RC algorithm to compute the scarred 
functions discussed in Sect.~\ref{sec:Intro} along different POs, and for different energy ranges.
For this purpose, we use the method introduced in Ref.~\cite{Fabio1}, that can be summarized as follows:
\begin{enumerate}
    \item Select a PO of the system.
    \item Prepare an initial \textit{tube} wavefunction $\psi_0^\text{tube}(x,y)$, localized on a tube
    along the PO, given by \cite{Fabio1}
    \begin{equation}
        \psi_0^\text{tube}(x,y) = \int_0^T e^{i \epsilon_n t/\hbar} \psi^\text{FG}(x,y,t) \; dt,
    \end{equation}
    where $T$ is the period of the periodic orbit, and $\epsilon_n$ is the associated Bohr-Sommerfeld (BS)
    quantized energy (see details below). 
    The function $\psi^\text{FG}(x,y,t)$ is a frozen Gaussian \cite{FrozenG} centered on the PO trajectory, 
    so that its probability density is forced to stay around such PO
    \begin{eqnarray}
        \psi^\text{FG}(x,y,t) = e^{-\alpha_x(x-x_t)^2 - \alpha_y (y-y_t)^2} \cdot \nonumber\\ 
        e^{i[p_{x_t}(x-x_t) + p_{y_t}(y-y_t) + \theta_t]/\hbar},
    \end{eqnarray}
    where $(x_t,y_y,p_{x_t},p_{y_t})$ is the trajectory of the PO, and $\alpha_x$ and $\alpha_y$ 
    are the widths along the two spatial dimensions. 
    In our case, we set $\alpha_x = \alpha_y = 1$.  
    The term $\theta_t$ is a phase defined as
    \begin{equation}
        \theta_t =  \frac{S(t)}{\hbar} - \mu \; \frac{\pi}{2}
    \end{equation}
    where $\mu$ is the Maslov index \cite{maslov}, ensuring that the wavefunction fulfils a 
    constructive interference along the PO. 
    The corresponding BS energies are computed by imposing that
    \begin{equation}
        \theta(T) = \frac{S(T)}{\hbar} - \frac{\mu \pi}{2} = 2\pi n 
        \label{eq:action}
    \end{equation}
    Using Eqs.~(\ref{eq:action}) and (\ref{eq:scaling}) it is found that
    \begin{equation}
        \epsilon_n = \left[\frac{2\pi \hbar}{S_0}\left(P\cdot n + \frac{\mu}{4} 
                      + \frac{P\cdot ND}{2}\right)\right]^{4/3},
    \end{equation}
    where $ND$ is the number of Dirichlet conditions on the wave functions at the symmetry lines 
    (axis and diagonals), $P$ is the ratio between the period of the full PO and that of the 
    desymmetrized PO (that is, reduced to the fundamental domain), and $n$ is the excitation number. 
    The values of these coefficients for the cases considered in this work are shown in Table~\ref{tab:PO}.
    \item  Compute the time evolution $\psi(x,y,t)$ under the Hamiltonian $\hat{H}(x,y)$. 
    Again, we use the FFT method to train a RC model and use it to further propagate 
    the wavefunction.
    \item Calculate the low-resolution energy spectrum $I(E)$. 
    First, we calculate the time-correlation function in Eq.~(\ref{eq:time-corr}). 
    Then, we obtain the energy spectrum $I(E)$ as the Fourier transform of the time-correlation function 
    \begin{equation}
        I(E) = \int_{-t_E}^{t_E} \; C(t) \; e^{-i E t/ \hbar} \; dt,
        \label{eq:low-spectrum}
    \end{equation}
    evaluated for a short period of time  $t_E$, taken as the Ehrenfest time
    \begin{equation}
        t_E = \frac{1}{2\lambda} \ln\left(\frac{\mathcal{A}}{\hbar}\right).
    \end{equation}
    In the case of our quartic potential $\lambda \approx  0.385 E^{1/4}$ is the Lyapunov exponent 
    of the system and $\mathcal{A} = 11.1 E^{3/4}$ is the area of a characteristic Poincaré surface 
    of section \cite{Fabio3}.  
    The peaks in this low-resolution spectrum $E_n$ will be used to compute the scarred functions.
    
    \item Calculate the scarred functions $\psi^\text{scar}(x,y)$ by performing the Fourier transform 
    of $\psi(x,y,t)$ at the energies $E_n$
    \begin{equation}
        \psi^\text{scar}_n(x,y) = \int_{-t_E}^{t_E} \; \psi(x,y,t) \; e^{i E_n t/\hbar} \; dt.
        \label{eq:initial}
    \end{equation}
\end{enumerate}

\begin{figure*}
\includegraphics[width=0.6\textwidth]{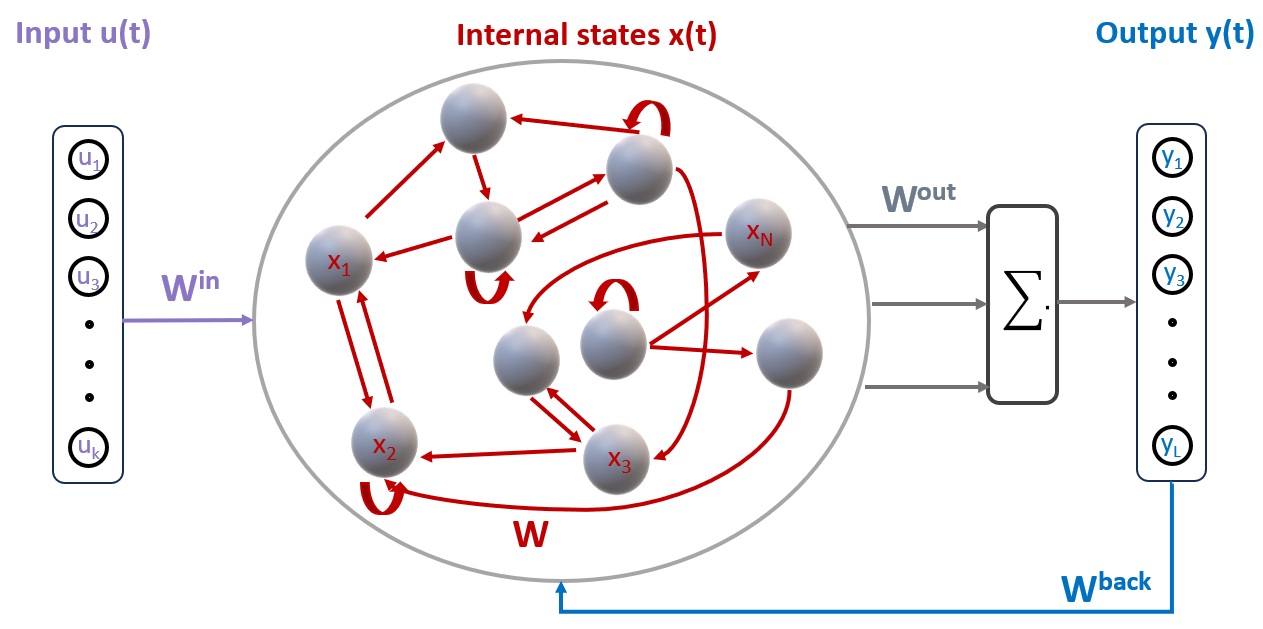}
\caption{Architecture of a RC model. 
Only the readout layer $W^{\text{out}}$ is learnt during training.}
\label{fig:RC}
\end{figure*}
\subsection{Reservoir computing}
\label{RC}
RC is a machine learning method that
has proven very useful in predicting the evolution of time
series. In the RC framework, the learning complexity of the algorithm is reduced to performing a linear regression. This machine learning algorithm consists of a neural network, where each of its nodes evolves in time with a dynamics that depends on the input-output dynamics. For this reason, the internal states of the network are called \textit{echo states}, 
since they can be thought of as an echo of their past \cite{ESN}. Then, a linear model is fit to find the relationship between the evolution of the internal states of the network and the actual dynamics of the system we aim to predict. The structure of the RC algorithm is schematically shown in Fig.~\ref{fig:RC},
where
\begin{itemize}
    \item $W^{\text{in}} \in M(\mathbb{R})_{N\times K}$ gives the weights from the input units to the internal states, 
    \item $W \in M(\mathbb{R})_{N\times N}$ gives the weights between the different internal states,
    \item $W^{\text{out}}\in M(\mathbb{R})_{L\times N}$ gives the wights from internal states to output units, and
    \item $W^{\text{back}}\in M(\mathbb{R})_{N\times L}$ gives the weights from the output units to the internal states,
\end{itemize}
and where
\begin{itemize}
    \item $\vec{u}(t)=(u_1(t),u_2(t),\ldots,u_K(t))$ is a $K$-dimensional vector giving the input units 
    at time $t$,
    \item $\vec{x}(t)=(x_1(t),x_2(t),\ldots,x_N(t)))$ is an $N$-dimensional vector giving the internal 
    states at time $t$, and
    \item $\vec{y}(t)=(y_1(t),y_2(t),\ldots,y_L(t)))$ is an $L$-dimensional vector giving the output 
    units at time $t$.
\end{itemize}
Notice that the matrices $W^{\text{in}}$, $W$ and $W^{\text{back}}$ are randomly chosen before the 
training phase so that they do not change during training. The only learnable parameters are the 
weights $W^{\text{out}}$. The steps to train the RC algorithm are the following:

\begin{enumerate}
    \item Generate the reservoir matrices $(W, W^{\text{in}}, W^{\text{back}})$ randomly. 
    \item Update the internal states by teacher forcing:
    \begin{eqnarray}
    \tilde{x}(t) &&= f(W^{\text{in}} u(t)+Wx(t-1)+W^{\text{back}}y_{\text{teach}}(t-1))\nonumber\\
    x(t) &&=(1 - \alpha) x(t-1) + \alpha \tilde{x}(t)
    \label{xn_train}
    \end{eqnarray}
    where $y_{\text{teach}}$ is the output that we want our network to predict, 
    $\alpha \in (0,1]$ is the leaking rate and $f$ is the activation function. 
    Usually, $f(\cdot)=\tanh(\cdot)$, which is applied component-wise.
    \item Discard a transient of $t_\text{min}$ states to guarantee the convergence of the reservoir dynamics.
    \item Find the readout matrix $W^\text{out}$ by minimizing the mean squared error with 
    $L^2$ regularization (ridge regression):
    \begin{eqnarray}
     MSE_r(y, y_{\text{teach}}) =&& \frac{1}{T - t_{\text{min}}} \sum_{t=t_{\min}}^T \Big(y_{\text{teach}}(t) \nonumber\\
     && - W^{\text{out}} x(t)\Big)^2 + \gamma ||W^{\text{out}}||^2,
     \label{ridge}
     \end{eqnarray}
    Notice that this step only requires performing a linear regression.
\end{enumerate}

The steps to make the predictions after training the network are the following:
\begin{enumerate}
    \item  Given an input $u(t)$, update the state of the reservoir:
    \begin{eqnarray}
        \tilde{x}(t) &&= f[(W^{\text{in}} u(t)+Wx(t-1)+W{^\text{back}}y(t-1)]\\ \nonumber
        x(t) &&=(1 - \alpha) x(t-1) + \alpha \tilde{x}(t)
        \label{xn_test}
    \end{eqnarray}
    where $y(t-1)$ is the prediction of the output at time $t-1$.
    \item Compute the prediction $y(t)$:
    \begin{equation}
        y(t) = W^{\text{out}} x(t).
    \end{equation}
\end{enumerate}

In our setting, the RC model is used without the input layer since
 we aim to propagate a dynamical system without using any explanatory variables. In this case, the term $W^{\text{in}}u(t)$ is removed from Eqs.~(\ref{xn_train}) and (\ref{xn_test}).

\subsection{Reservoir computing for quantum systems}
\label{sec:RC_QuantumSystems}

The propagation of the initial wavepacket is done by training a RC model with the wavefunctions 
calculated by the FFT method until time $t_\text{train}$ 
and then using the RC model to predict the evolution until time $t_\text{train} + t_\text{test}$. 
In the RC framework, $\psi(x,y,t)$ can be represented as a set of matrices $\{\psi(x,y,t)\}_t$, 
where each matrix contains the values of $\psi(x,y)$ at time $t$ in a grid of points spanning $(x,y)$. 
However, in Ref.~\cite{Domingo} we proposed a more efficient method to adapt RC to propagate 
quantum systems. 

Two main challenges need to be overcome to adapt the traditional RC to propagate wavefunctions
\begin{itemize}
    \item \textbf{Complex numbers}:  In the usual RC framework all the vectors and matrices are real-valued. However, the target time series for this quantum problem is a wavefunction $\psi(x,y,t)$ which, in general, takes complex values. We adapted the RC algorithm to work with complex numbers by proposing the activation function
    \begin{equation}
    f(x) = \tanh(\Re(x)) + i \; \tanh(\Im(x))
    \end{equation}
    and providing a closed-form solution for the complex-valued ridge regression
    \begin{equation}
    W^\text{out} = \Big(X^* X + \alpha \mathbb{I}\Big)^{-1} \times \Big(X^*f_\text{out}^{-1}(\vec{y}_\text{teach})\Big),
    \end{equation}
    where $^*$ denotes the conjugate transpose, and $X$ is the matrix containing the internal states. 
    \item \textbf{High-dimensional data}: The target data is represented as a matrix, where each entry is the value of the wavefunction 
    in a discretized spatial grid.  The size of this matrix increases exponentially with the dimension of the physical system, that is, the dimension of $(x,y)$. With large input sizes we need to use large reservoirs $W$. If the complexity of the neural network is too large, the linear model can easily overfit the training data, and therefore be unable to predict the evolution of the dynamical system. For this reason, we proposed a new training strategy that reduces the overfitting of the algorithm. This learning strategy shows the reservoir how small predicting errors modify the internal states during the test phase. This reduces overfitting, decreasing the test error significantly. For further information on this method, refer to Ref. \cite{Domingo}. 
\end{itemize}

Table~\ref{tab:params} shows the training parameters used for each of the systems studied in this work. 
For all systems we used $f^\text{out} = \mathbb{I}$ and $t_\text{min} = 500$. 
For the multi-step process described in Ref.~\cite{Domingo}, we used 80\% of the training data for 
the first training step and 20\% for the second step. 
The spectral radius of the internal states $W$ is set $\rho(W) = 0.5$, and the density of $W$ is set to 0.005.
Even though a cumbersome machine learning model such as Bayesian optimization could have been used 
to select the hyperparameters, we consider that, in this case, such methods are not necessary. 
We follow instead the criteria given in Ref.~\cite{TrainingParams} to choose the appropriate training 
parameters.
    
\begin{table}
    \centering
    \begin{tabular}{lcccc}
    \hline \hline 
         Calculation               & $\alpha$ & $\gamma$ & $N$  \\  \hline  
         Eigenfunctions $E=1$      & 0.2   &  0.001 & 2000 \\
         Eigenfunctions $E=10$     & 0.2   &  0.001 & 2000 \\ 
         Eigenfunctions $E=100$    & 0.017 &  0.1   & 10000 \\
         Horizontal/vertical scar  & 0.1   &  1.0   & 1500 \\
         Quadruple-loop scar       & 0.3   &  0.1   & 2000 \\
         Bowtie hor/ver scar       & 0.3   &  0.1   & 3000 \\
         Square scar               & 0.3   &  0.1   & 3000 \\   \hline 
    \end{tabular}
    \caption{Reservoir computing training parameters, as defined in Sect.~\ref{RC}, used in the calculation
    of different eigenfunctions and scars for the quartic potential in Eq.~(\ref{eq:quartic}).}
    \label{tab:params}
\end{table}    

%
\section{Results}
\label{sec:Results}
%
\begin{figure}[!b]
 \includegraphics[width=1.0\columnwidth]{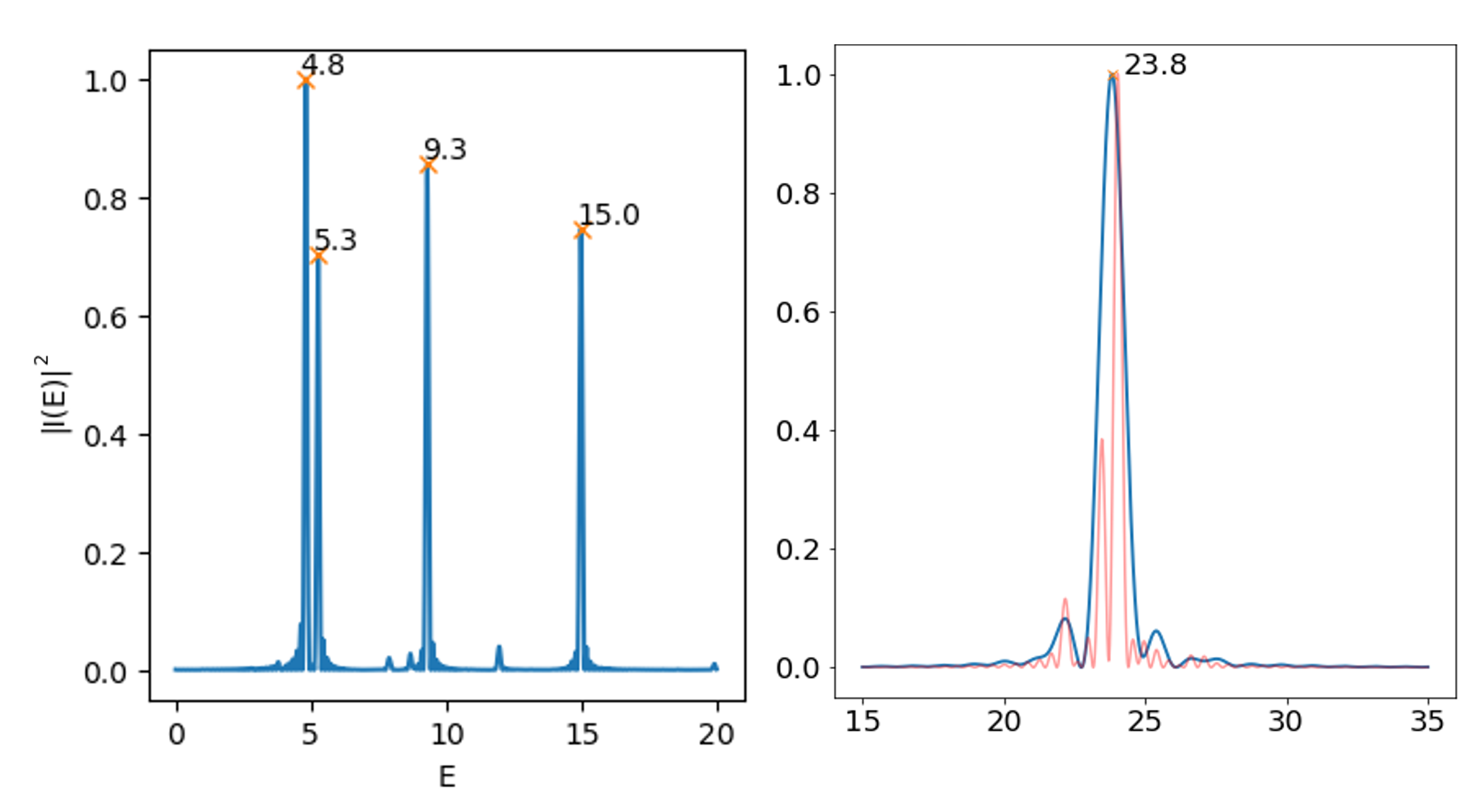}
  \caption{(Left) Example of the modulus of the energy spectrum $|I(E)|^2$ for the system at $E=10$. 
  (Right) Long time energy spectrum $|I(E)|^2$, in red (light gray), and its low-resolution version, 
  in blue (dark gray), 
  for the quadruple-loop scar function at energy $E = 22.824$.}
 \label{fig:spectrum}
\end{figure}
\begin{table}
    \centering
    \begin{tabular}{rcccc}
    \hline
    \hline
         $E_0$  & $E_n$   & $E_n$ MSE           & $\phi_n$ MSE  \\
        \hline 
        $1$     & .56323  & $4 \cdot 10^{-10}$   & $5 \cdot 10^{-7}$\\
                & 1.8848  & $< 10^{-10}$        & $4 \cdot 10^{-7}$\\
                & 2.8638  & $ 4 \cdot 10^{-10}$ & $1 \cdot 10^{-5}$\\ \hline
        $10$    & 4.8286  & $ 1 \cdot 10^{-8}$ & $6 \cdot 10^{-6}$\\
                & 5.2584  & $ 1 \cdot10^{-8}$  & $7 \cdot 10^{-6}$\\
                & 9.3067  & $ 1 \cdot 10^{-8}$ & $9 \cdot 10^{-6}$\\
                & 14.9547 & $ 1 \cdot 10^{-8}$ & $1 \cdot 10^{-5}$\\ \hline
        $100$   & 28.6843 & $ 4 \cdot 10^{-8}$ & $8 \cdot 10^{-6}$\\
                & 46.8794 & $ 1 \cdot 10^{-6}$ & $2 \cdot 10^{-5}$\\
                & 48.5846 & $ 6 \cdot 10^{-6}$ & $3 \cdot 10^{-6}$\\
                & 50.2605 & $ 6 \cdot 10^{-7}$ & $8 \cdot 10^{-7}$\\
                & 52.5642 & $ 3 \cdot 10^{-5}$ & $4 \cdot 10^{-6}$\\
                & 56.9221 & $ 3 \cdot 10^{-6}$ & $5 \cdot 10^{-7}$\\ 
                & 83.1969 & $ 1 \cdot 10^{-5}$ & $6 \cdot 10^{-7}$\\
                & 87.9381 & $ 9 \cdot 10^{-8}$ & $ 6 \cdot 10^{-7}$\\
                & 89.3304 & $ 2 \cdot 10^{-5}$ & $ 9 \cdot 10^{-6}$\\
        \hline
    \end{tabular}
    \caption{Eigenenergies $E_n$ for the different eigenfunctions $\phi_n$ shown in 
    Figs.~\ref{fig:1}, \ref{fig:2}, and \ref{fig:3}, together with the corresponding mean squared 
    errors calculated by comparison with the `exact' variational ones, obtained from the three
    initial wavepackets specified in Table~\ref{tab:initial_waves}.}
    \label{tab:mse}
\end{table}
%
\begin{figure}
 \includegraphics[width=1.0\columnwidth]{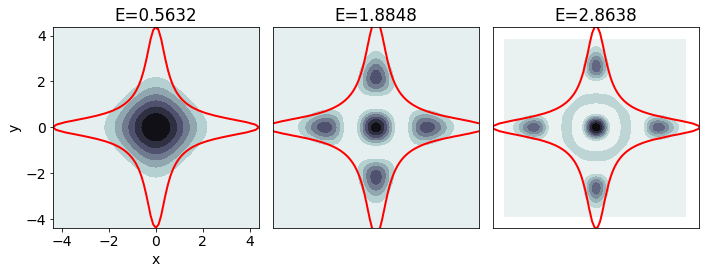}
  \caption{Probability density of the eigenfunctions of the coupled quartic oscillator obtained from 
  propagating an initial wavepacket with energy $E=1$. 
  The red (light grey) solid lines depict the equipotential curves.}
 \label{fig:1}
\end{figure}
%
\begin{figure}[!b]
 \includegraphics[width=0.85\columnwidth]{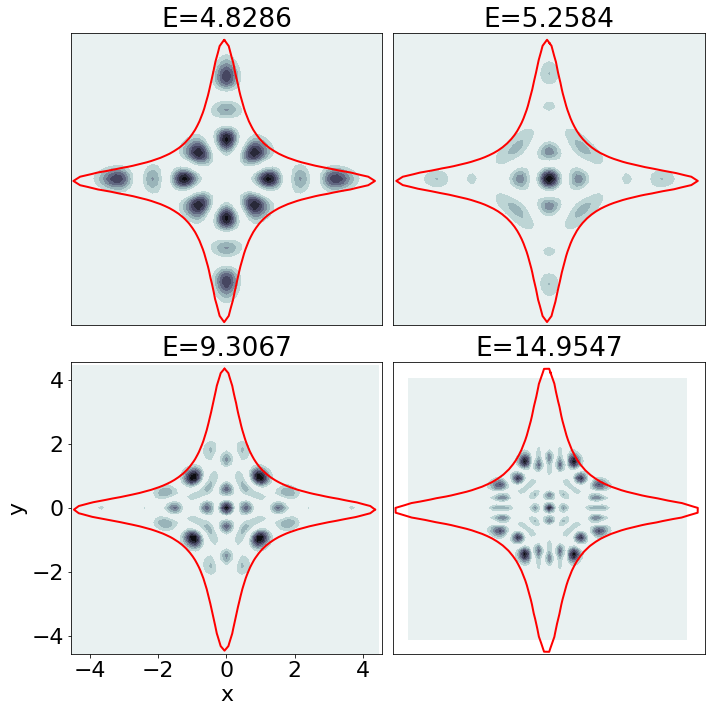}
  \caption{Same as Fig.~\ref{fig:1} for $E=10$, scaled to the domain corresponding to $E=1$.}
 \label{fig:2}
\end{figure}
%
\begin{figure*}
 \includegraphics[width=0.60\textwidth]{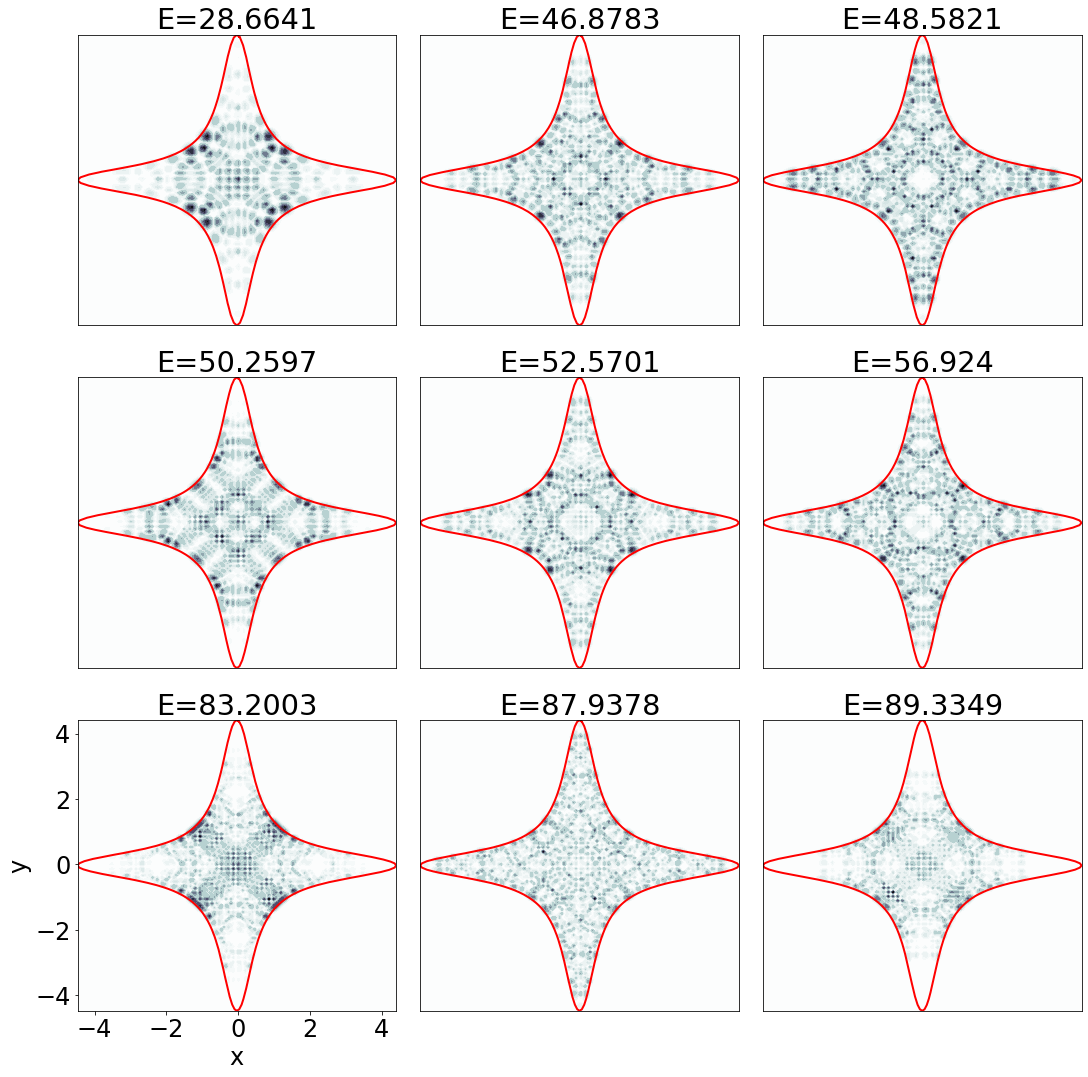}
  \caption{Same as Fig.~\ref{fig:2} for $E=100$.}
 \label{fig:3}
\end{figure*}
\begin{table}[t!]
    \centering
    \begin{tabular}{lccc}
    \hline \hline
         System                   & Method &  Execution \\
                                  &        &   time (in hours) \\
    \hline
        Eigenfunctions $E=1$      & FFT & 3:35 \\
                                  & RC  & 0:10 \\
        Eigenfunctions $E=10$     & FFT & 4:50 \\
                                  & RC  & 0:36 \\
        Eigenfunctions $E=100$    & FFT & 24:47 \\
                                  & RC  & 1:34 \\    
        Horizontal/vertical scar  & FFT & 2:58 \\
                                  & RC  & 0:12 \\   
        Quadruple-loop scar       & FFT & 1:32 \\
                                  & RC  & 0:07 \\ 
        Bowtie hor/ver scar       & FFT & 3:27 \\
                                  & RC  & 0:24 \\  
        Square scar               & FFT & 3:44 \\
                                  & RC  & 0:25 \\   
    \hline
    \end{tabular}
    \caption{Execution times for the Fast Fourier Transform integrator \cite{kosloff}
    and the RC model, for all cases studied in this work.}
    \label{tab:time}
\end{table}

In this section, we use our RC method to calculate eigenfunctions and scarred functions of the coupled 
quartic oscillator (\ref{eq:quartic}), and examine its performance for these two tasks.
Let us start with the calculation of eigenfunctions.

First, we compute the quantum dynamics generated by three different initial wavepackets at $E_0=1,10$ and $100$,
respectively, and the parameters reported in Table~\ref{tab:initial_waves}, and from them we compute
the corresponding spectra and eigenstates, with the method described in Sect.~\ref{sec:methods}.

In the left panel of Fig.~\ref{fig:spectrum} we present, as an example, the spectrum for the 
$E_0=10$ case. The packet is launched along the main diagonal, and accordingly, only totally
symmetric eigenstates, i.e.~belonging to the $A_1$ symmetry class, are generated.
As can be seen, four conspicuous peaks, at the eigenenergies reported in Table~\ref{tab:mse}, 
are clearly observed.

The associated eigenfunctions are shown in Figs.~\ref{fig:1}, \ref{fig:2} and \ref{fig:3}, respectively. 
For comparison, we have also calculated the eigenenergies and eigenfunctions using the variational method \cite{variational_original}, which will be considered to be the \textit{exact} ones, in contrast to the \textit{predicted} ones produced with RC. 
As can be seen the first three $A_1$ eigenstates (see Fig.~\ref{fig:1}) corresponds, as expected,
to states $(n_x,n_y)=(0,0), (1,1)$, and $(2,2)$, respectively.
In the next two figures, we show more excited eigenfunctions, which then exhibit more complicated
nodal pattern topologies. 
Interesting to note that some very excited cases, in particular those for 
$E_n=50.2597, 83.2003$ and $89.3349$ in Fig.~\ref{fig:3}, appear as clear scars in the original sense 
of Heller \cite{Heller} of the double-loop and diagonal POs.

The mean squared error (MSE) of the predicted eigenenergies and eigenfunctions, compared to the exact ones, 
are provided in Table~\ref{tab:mse}.  
As can be seen, the MSE of the eigenenergies increases with the energy of the system, 
since the eigenfunctions become more complex and thus harder for the RC algorithm to predict 
its time evolution.
However, the MSE of all the energies and eigenfunctions is of the order or smaller than
$1\times10^{-5}$, which means that the machine learning RC model can correctly propagate in time
the quantum states of this chaotic system. 
Therefore, the RC algorithm, being agnostic to the underlying physical model of the system, 
can reproduce the dynamics of complex quantum chaotic systems, 
which proves the versatility of the algorithm to be adapted to multiple quantum systems \cite{Domingo}.

The advantage of using a machine learning algorithm instead of a classical numerical integrator 
is the execution time needed to propagate the initial state in time. 
Table~\ref{tab:time} shows the execution time for the FFT integrator and the RC models. 
These results show that the RC method is much faster than the classical integrator, 
which makes it suitable for predicting the long-time evolution of quantum states.
\begin{figure*}
 \includegraphics[width=0.60\textwidth]{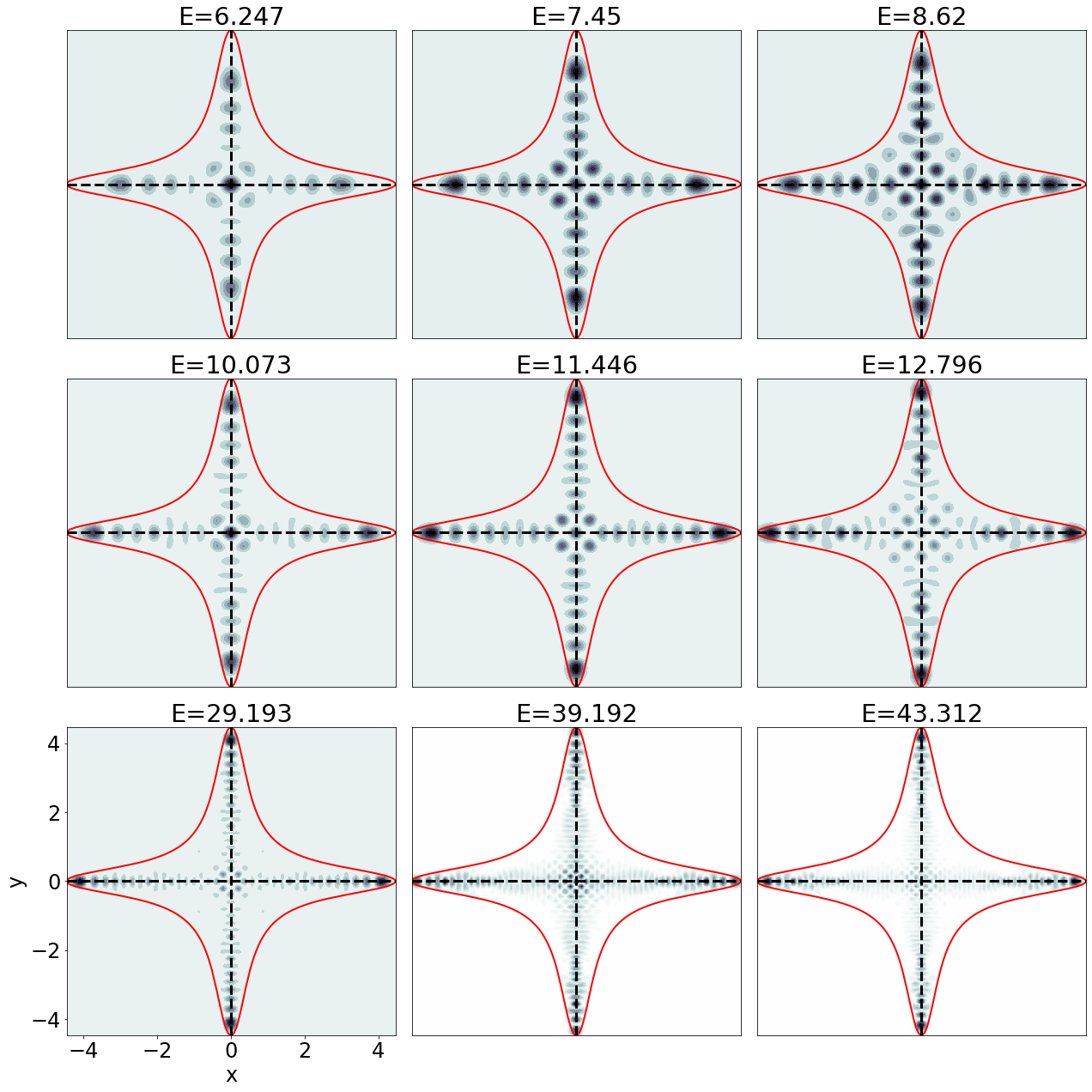}
  \caption{Probability density of the scar function for the horizontal/vertical periodic orbit, 
  at different energies (see Table \ref{tab:PO}). 
  The solid red (light gray) lines depict the equipotential curves, 
  while the dashed black lines illustrate the periodic orbits}.
 \label{fig:4}
\end{figure*}
%
\begin{figure*}
 \includegraphics[width=0.75\textwidth]{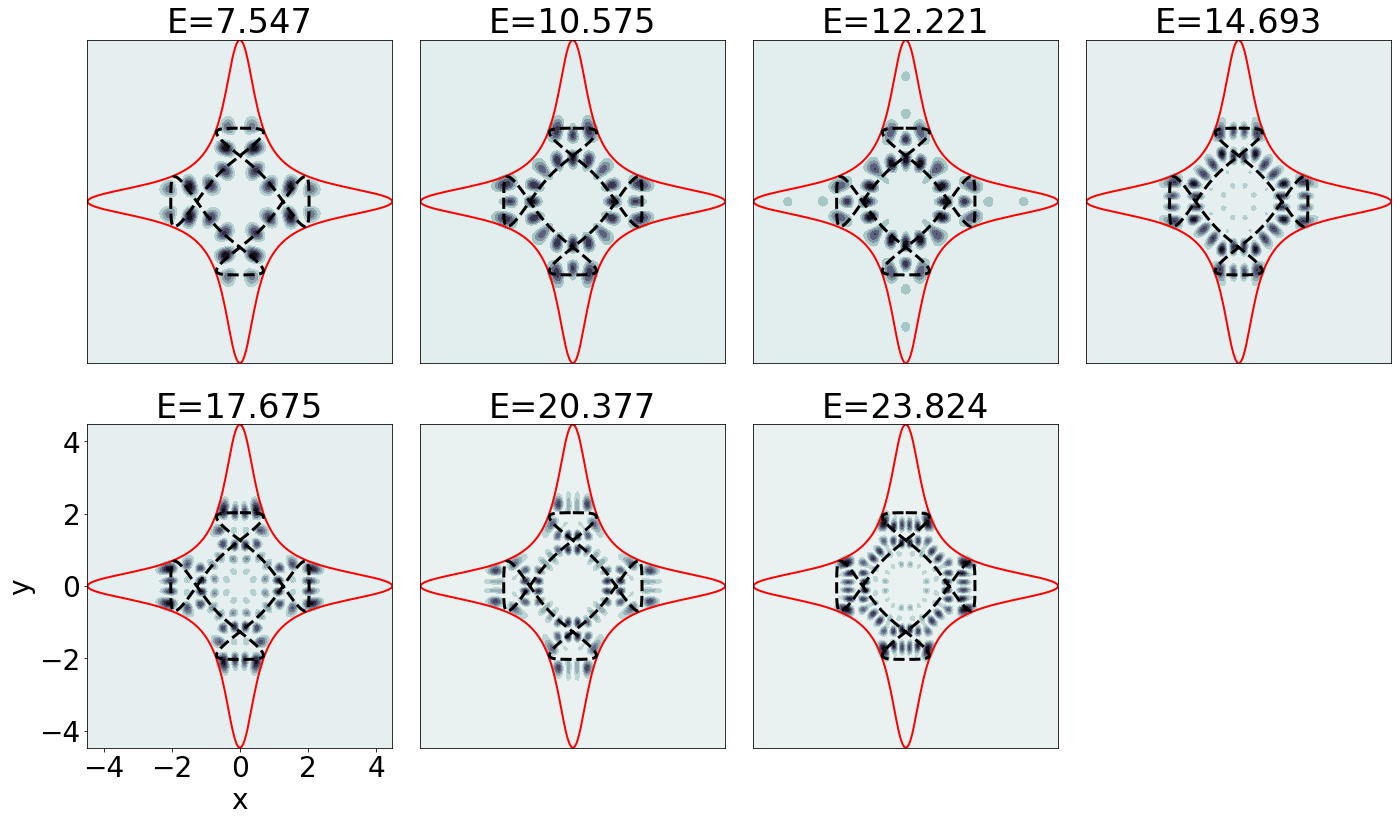}
  \caption{Same as Fig.~\ref{fig:4} for the quadruple-loop periodic orbit.}
 \label{fig:5}
\end{figure*}
%
\begin{figure*}
 \includegraphics[width=0.75\textwidth]{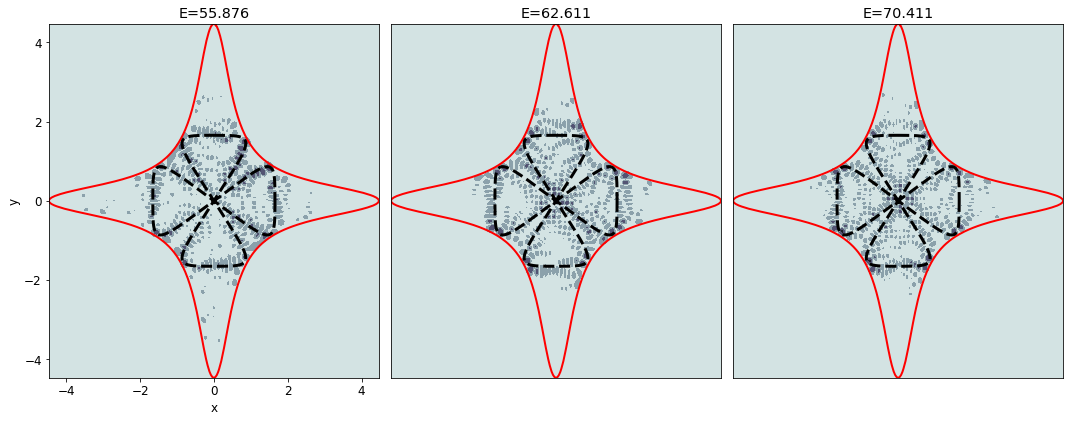}
  \caption{Same as Fig.~\ref{fig:4} for the bowtie periodic orbit.  }
 \label{fig:7}
\end{figure*}
\begin{figure*}
 \includegraphics[width=0.65\textwidth]{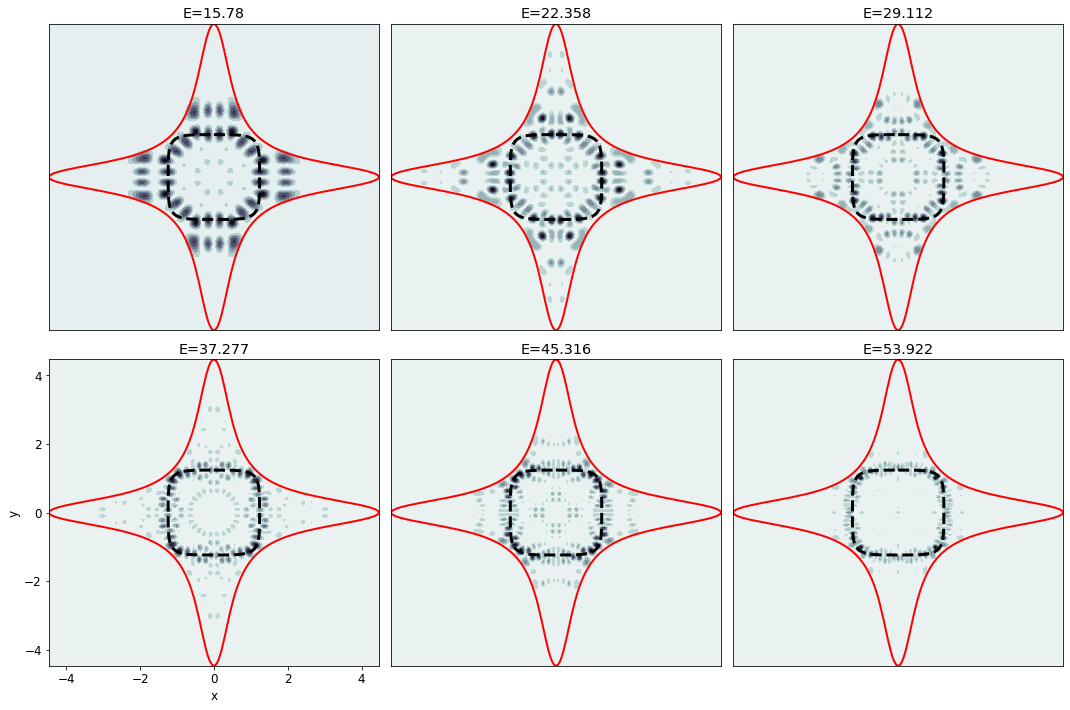}
  \caption{Same as Fig.~\ref{fig:4} for the square periodic orbit.}
 \label{fig:6}
\end{figure*}

Apart from the eigenfunctions, we have also used RC to calculate some scarred functions 
at different energies for the four POs shown in Fig.~\ref{fig:po}. 
Notice that the horizontal/vertical and bowtie horizontal/vertical POs are in fact made of two POs 
(solid and dashed lines in Fig.~\ref{fig:po}) to take into account the symmetry of 
Hamiltonian (\ref{eq:quartic}). 
Similarly to what was done in the case of the eigenstates, this choice is made so that
the resulting scarred functions belong to the $A_1$ irreducible representation, that is, 
they are symmetric with respect to the $x$, $y$ axis, and the two diagonals $y = \pm x$. 

\begin{figure*}
 \includegraphics[width=0.85\linewidth]{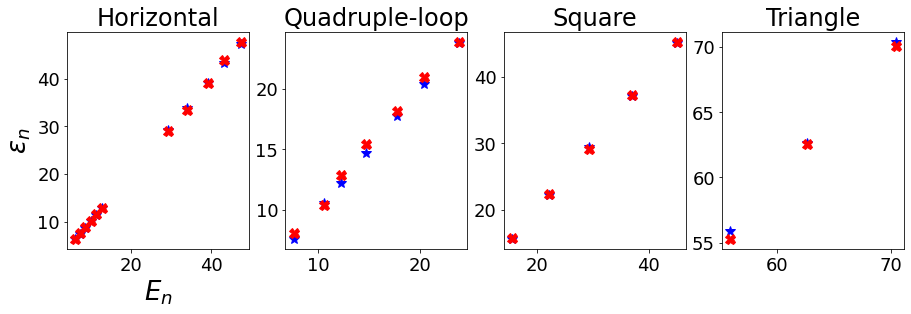}
  \caption{Comparison of the semiclassical energies (red crosses) with the scar energies obtained with the reservoir computing approach (blue stars).}
 \label{fig:8}
\end{figure*}

As explained in Sect.~\ref{sec:scarfunctions}, the energies of the scarred functions are obtained 
by finding the peaks of the low-resolution energy spectrum generated from a wavepacket launched 
along the selected PO. 
An example of the corresponding high (in red) and low-resolution, resolved up to the Ehrenfest time,
(in blue) spectra for the quadrupe-loop PO at $E = 22.824$ are shown in the right panel of
Fig.~\ref{fig:spectrum}. 
As can be seen, for the longer time the energy peaks in the spectrum are more pronounced and
better resolved, while the low-resolution spectrum has wider peaks which contain several high-resolution 
peaks. 
This is an effect due to the inclusion or not of dynamical information beyond the first recurrence
along the PO, and was thoroughly discussed in Refs.~\cite{Wis1,Wis2}.

Once the energies of the scar functions have been calculated, the associated scarred functions 
are obtained with the procedure described in Sect.~\ref{sec:scarfunctions}.
Figures~\ref{fig:4}, \ref{fig:5}, \ref{fig:7}, and \ref{fig:6} show some scarred functions calculated
for the four POs in Fig.~\ref{fig:po} at different energies 
(scaled to energy $E=1$ for an easier visualization). 
We see that all the scarred functions appear very well localized around their associated PO, 
showing an increasing consecutive excitation as energy increases.
Actually, in almost all cases scar quantum numbers can be easily assigned, being this specially clear 
in all panels of Figs.~\ref{fig:4} and~\ref{fig:5}.
On the other hand, in some cases an 'interaction' between POs is apparent. 
For example, the scarred function shown in Fig.~\ref{fig:6} for $E=15.78$ is mainly localized 
around the square PO, but it also has visible and easily countable nodes along the quadruple-loop PO. 
This is because the time-propagation of the initial wavefunction gathers information 
about the phase space around both POs, which then shows in the resulting scar. 
Moreover, notice that in Fig.~\ref{fig:5}, there is a quadruple-loop scar function at a similar energy,
i.e.~$E=14.693$, to that of the square scarred function mentioned before, which is an indication of
the possibility of interaction between them. 
On the other hand, other scarred functions appear totally localized around one single PO 
(see, for example, the horizontal scarred functions in Fig.~\ref{fig:4}, or the square scarred function 
at $E=53.922$ in Fig.~\ref{fig:6}).
Again, a complete discussion of this effect can be found in Refs.~\cite{Wis1,Wis2}.

To end this subsection, we compare the energies obtained by our RC method with the Bohr-Sommerfeld 
quantized energies. The result is shown in Fig.~\ref{fig:8}. 
As can be seen, the energies obtained with both methods are very similar. 
Some small differences between the Bohr-Sommerfeld energies and the quantum energies obtained with 
the RC method may, however, be expected. 
First, the Bohr-Sommerfeld energies are a semiclassical approximation of the true quantum energies,
which include all quantum effects. 
Second, the Bohr-Sommerfeld approximation only takes into account the influence of a single PO
for each scarred state, and we have already discussed that some of the resulting scarred functions 
obtained with the RC method are the result of (small) interactions between different POs, 
which also causes small differences in the obtained energies.

 In conclusion, our results show that the RC method is able to predict the time evolution of the 
 coupled quartic oscillator for different initial conditions, and obtain, with high accuracy, 
 the associated eigenenegies, eigenfunctions and scarred functions.
 This efficient machine learning algorithm is faster than the traditional PDE solver, 
 and it is easily adaptable to different systems.

\section{Conclusions} 
\label{sec:Conclusions}
In this paper, we introduced a machine learning method to efficiently calculate the eigenstates 
and scarred functions of quantum systems. 
In particular, the novel algorithm called RC, which has been shown to excel at predicting 
the evolution of chaotic time series \cite{chaos1, chaos2}, is used. 
As an illustration, this algorithm has been applied to the coupled quartic oscillator,
which has been extensively studied in the field of quantum chaos. 
From a suitable initial wavepacket, we have used RC to propagate the quantum state in time.
Then, by performing a Fourier transform we have been able to recover the eigenstates and eigenenergies 
of the system at different energies in the vicinity of the mean energy of the packet. 
Moreover, given an initial condition localised around a PO, we have calculated the associated
scarred functions at multiple energies by computing the low-resolution spectrum, for four different 
POs of the quartic oscillator. 
The possibility of performing extensive calculations involving many excited scarred states is
of importance in the field of quantum chaos \cite{Wis3,Wis4,Hummel}.

The results have been compared with more traditional methods, such as the variational method 
and the Bohr-Sommerfeld semiclassical quantization.  
We have seen that even with high energies and complex wavefunction, the RC algorithm can 
reproduce the eigenstates and scarred functions with high accuracy, meaning that the method can 
correctly propagate the initial wavefunction in time. 
Moreover, the simple training strategy of RC provides fast execution times when compared to the other 
classical methods. 
Finally, as we have shown in a previous work \cite{Domingo}, RC is easily adaptable to propagate 
different systems, which makes the method very versatile.

\section*{Code Availability}
The code and data that support the findings of this study are openly available in \href{https://github.com/laiadc/Scars\_RC}{https://github.com/laiadc/Scars\_RC}.

\section*{Acknowledgments}
The project that gave rise to these results received the support of a fellowship from "la Caixa" Foundation 
(ID 100010434). The fellowship code is LCF/BQ/DR20/11790028.
This work has also been partially supported by the Spanish Ministry of Science, Innovation and Universities, 
Gobierno de Espa\~na, under Contract No.~PID2021-122711NB-C21.
\bibliography{bibliography}
\end{document}